\documentclass[aps,pra,superscriptaddress,twocolumn,10pt,nofootinbib,preprintnumbers,showdata]{revtex4-2}
\usepackage{bm}
\usepackage{epsfig}
\usepackage{graphics}
\usepackage{amsmath}
\usepackage{xcolor}

\begin{document}

\title{Collective ion dynamics in Coulomb one-component plasmas \\  within the self-consistent relaxation theory}

\author{ \firstname{Ilnaz~I.}~\surname{Fairushin}}
\author{ \firstname{Anatolii~V.}~\surname{Mokshin}}

\affiliation{Department of Computational Physics, Institute of Physics, Kazan Federal University, 420008 Kazan, Russia}

\begin{abstract}
In this paper, we present the theoretical formalism describing the collective ion dynamics of the nonideal Coulomb classical one-component plasmas on the basis of the self-consistent relaxation theory. The theory is adapted to account for correlations between the frequency relaxation parameters that characterize the three- and four-particle dynamics and the parameters associated with the two-particle dynamics. The dynamic structure factor spectra and dispersion characteristics calculated for a wide range of wave numbers are in agreement with the molecular dynamics simulation data and the results obtained with the theory of the frequency moments. The proposed formalism reproduces all the features inherent to the Coulomb one-component plasmas and requires only knowledge of the coupling parameter and the information about the structure.
\end{abstract}

\maketitle

\section{Introduction}
The Coulomb one-component plasmas (COCP) is the specific system of identical charged point particles in a uniform neutralizing background \cite{Brush, Hansen2, Baus, Ichimaru, MithenRC, MithenPRE2012, DaligaultPRL, Takeno, Fortov3, Potekhin, Baiko, Dornheim, Arkhipov, TkachenkoPRE2020, Murillo_2019}. Together with the hard sphere model, the COCP occupies an important place in the theory of simple liquids and plays a key role in the physics of extreme states of matter \cite{Takeno, Fortov3}. In nature, the COCP is realized in such objects as neutron star crusts, interiors of white dwarfs and giant planets \cite{Takeno, Fortov3, Potekhin, Baiko, Dornheim}. Point particles of the COCP interact with each other through the Coulomb potential
\begin{equation}
u(r) = \frac{(Ze)^2}{4\pi\varepsilon_0r}\,,
\label{coulomb_pot}
\end{equation}
where $r$ is the distance between particles, $\varepsilon_0$ is the electrical constant, $Z$ is the particle charge in units of electron charge $e$. Despite the simple analytical form of this potential, the COCP is nontrivial system. First of all, the Coulomb interaction potential is long ranged. Unlike, for example, the Lennard-Jones potential, the Coulomb potential (as well as the Yukawa potential)  is purely repulsive. Therefore, laboratory and natural implementations of such systems are always accompanied by the presence of external factors that stabilize them and keep the system in equilibrium. For example, in experiments with dust particles, this is usually an external electric or magnetic field, which creates the so-called trap that holds particles within a finite volume \cite{Fortov3}. When modeling by the molecular dynamics (MD) method, the stability of these systems can be achieved by using periodic boundary conditions. In this case, however, the use of the standard method of taking into account the interaction with the nearest neighbours can give correct results only for intermediate and strongly screening Yukawa systems. The Coulomb systems as well as the Yukawa systems with a large screening length require considering the long-range effect of the corresponding potentials. As a rule, this is achieved using the modified Ewald summation method or its analogs, in which the effective interaction potential of particles (charges) takes into account their interaction with the background of the opposite sign \cite{Brush, Salin}. Thus, when describing the physical properties of the Coulomb system, it is always necessary to take into account the presence of the uniform neutralizing background of opposite sign. For example, in a general case, if particles of some system interact through some potential $\phi(r)$ and their mutual arrangement is reproduced by the radial distribution function $g(r)$, then the reduced excess internal energy per particle is defined by the following relation:
\begin{equation}
U_{ex} = \frac{2\pi\rho}{k_BT} \int_{0}^\infty \phi(r)g(r)r^2dr. \nonumber
\label{U_ex}
\end{equation}
Here, $\rho$ is the number density of particles, $k_B$ is the Boltzmann constant, and $T$ is the absolute temperature. Then, in the case of the Coulomb potential, $\phi(r)=u(r)$, the last equation will produce infinite energy. To calculate $U_{ex}$ in the case of the COCP, it is necessary to replace $g(r)$ with $g(r)-1$, which actually means taking into account the presence of a uniform neutralizing background. On the other hand, the interaction of the COCP charges with the opposite background appears at local deviations of their number density from some average value $\rho$. This interaction is characterized by forces whose amplitudes are directly proportional to the amplitude of the displacement of the point charges relative to the background \cite{Brush, Demyanov}. This leads to the appearance of a collective oscillatory motion of the COCP particles with the certain inherent frequency $\omega_p$. The frequency $\omega_p$ is called the plasma frequency and is defined as follows \cite{Brush, Hansen2, Baus, Ichimaru, MithenRC, MithenPRE2012, DaligaultPRL, Takeno, Fortov3}:
\begin{equation}
\omega_p = \sqrt{\frac{(Ze)^2\rho}{\varepsilon_0m}}.
\label{wp}
\end{equation}
Here, $m$ is the mass of the particles. On the other hand, as we know, the frequency of natural oscillations of a spring pendulum is determined by a relation of the form:
\begin{equation}
\omega_K = \sqrt{\frac{K}{M}},
\label{wK}
\end{equation}
where $K$ is the elasticity coefficient of the spring pendulum and $M$ is the mass of the pendulum. Comparing Eqs. \eqref{wp} and \eqref{wK}, one can see that the quantity $(Ze)^2\rho/\varepsilon_0$ actually represents the effective elasticity coefficient of the COCP.

The characteristic frequency of collective vibrations of particles at finite spatial scales will differ from $\omega_p$, i.e., there is a dispersion dependence of the frequency $\omega$ on the wave number $k$. Full information about the vibrational processes associated with density redistribution at different spatial scales of an equilibrium multiparticle system can be obtained from such quantity as the dynamic structure factor $S(k, \omega)$ \cite{Balucani}. The physical meaning of the dynamic structure factor can be interpreted as the intensity of fluctuations of particle number density in the system with different frequencies $\omega$ on various spatial scales $L\sim2\pi/k$. In the case of simple single-component liquids, the spectrum $S(k, \omega)$ at the fixed $k$ is characterized by one central and two symmetric side peaks \cite{Balucani}. The central peak corresponds to nonpropagating isobaric entropy fluctuations, and side peaks correspond to adiabatic pressure fluctuations, which propagate in space. The positions of these peaks on the frequency $\omega$ axis as well as their widths are associated with values of thermal diffusivity, attenuation coefficient, and sound velocity for the system.

The specificity  of the Coulomb system is largely due to the long-range interaction of particles (charges) and directly appears also in the form of the spectra $S(k, \omega)$ mainly in the long-wave limit (i.e., at $k\rightarrow0$). First, in the case of the COCP, the central (Rayleigh) peak is practically absent \cite{Baus, TkachenkoPRE2020}. This is due to the fact that the resulting arbitrary local redistribution of the particle number density rapidly propagates throughout the Coulomb system without appreciable transfer of thermal energy. Second, the side peaks of $S(k,\omega)$ spectra are located near the frequency $\omega_p$ and tend to it in low-$k$ limit ($k\rightarrow0$). This means that the $k$ dependence of the side peak positions of $S(k, \omega)$ spectra, which is denoted as $\omega_c(k)$, has a frequency (energy) gap at $k=0$ and the width of this gap is 
\begin{equation}
\omega_p=\lim_{k\rightarrow 0}\omega_c(k).
\end{equation}
Note that in the case of systems with a short-range interparticle interaction potential, the dispersion dependence $\omega_c(k)$ at $k\rightarrow0$ is linear: $\lim_{k\rightarrow0}\omega_c(k)=c_sk$, where $c_s$ is the speed of sound.  The shift of the side peak position as $k$ increases into the region of higher or lower frequencies compared to the frequency $\omega_p$ depends on the thermodynamic state of the system, which is determined by a single quantity -- the so-called coupling parameter
\begin{equation}
\Gamma = \frac{(Ze)^2}{4\pi\varepsilon_0ak_BT}\,.
\label{gamma}
\end{equation}
Here, $a = (3/4\pi \rho)^{1/3}$ is the radius of the Wigner-Seitz cell. The coupling parameter $\Gamma$ is approximately the ratio of the potential energy between two particles to the average thermal energy of a particle. The greater the value of the parameter $\Gamma$, the more coupled the system is. Thus, at $\Gamma\lesssim175$, the COCP is a disordered system, while, at $\Gamma\gtrsim175$, it is a crystal with the bcc lattice \cite{Hansen2, Baus}. When the thermal energy of motion of the particles is much larger than their interaction energy, i.e., at $\Gamma\ll 1$, ideal Coulomb gas is realized. In this case, the $k$ dependence of the characteristic frequency of collective excitations is given by the well-known Bohm-Gross dispersion relation \cite{Kampen, BohmGross}: 
\begin{equation}\label{BGdisp}
\omega^{(BG)}(k) = \omega_p\sqrt{1+\frac{(ka)^2}{\Gamma}}.
\end{equation}

\begin{figure}[h]
%	\centering
\includegraphics[width=7.8cm]{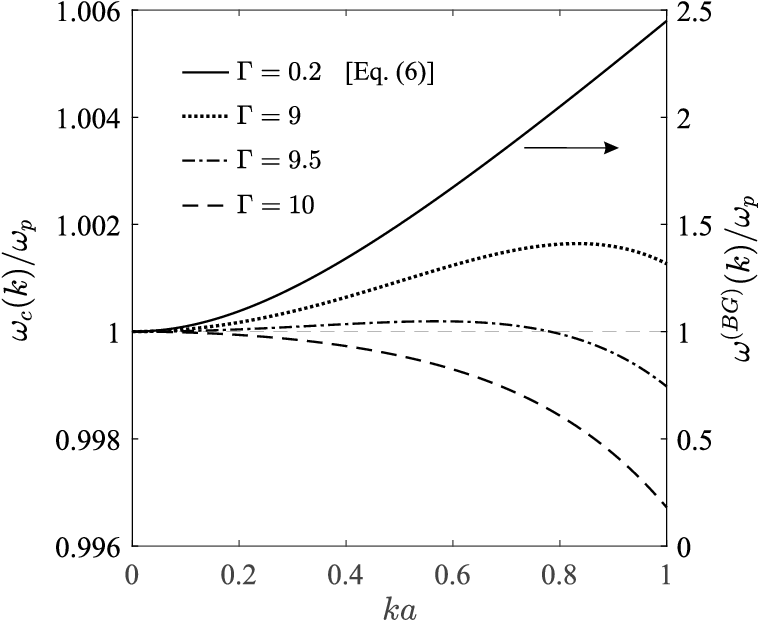}
\caption{Dispersions of side peak in $S(k, \omega)$ spectra for the COCP (reproduced from \cite{MithenAIP}) at different coupling parameters $\Gamma$. Solid line represents the Bohm-Gross dispersion relation (\ref{BGdisp}) at $\Gamma=0.2$.}\label{Disp_Mithen}
\end{figure}

For a strongly coupled COCP at $\Gamma$ values exceeding some critical value $\Gamma_c$, the side peak of $S(k, \omega)$ spectra at low-$k$ range with increasing $k$ shifts to a lower frequency compared to the plasma frequency $\omega_p$ \cite{HansenJPL, MithenAIP, KhrapakDisp}.
In this case, the so-called negative dispersion mode is realized and the dispersion dependence $\omega_c(k)$ will satisfy the condition:
\begin{equation}
\frac{d\omega_c(k)}{dk}<0 \,\,\, \text{at} \,\,\, k\rightarrow 0.
\end{equation}
The results of molecular dynamic (MD) simulations of the COCP \cite{MithenAIP} reveal that the critical value of the coupling parameter is $\Gamma_c \approx 9. 5$  (Fig. \ref{Disp_Mithen}). It is noteworthy that the $k$ dependence of $\omega^{(BG)}(k)$ is similar to the dependence of the total energy of mass particles on their momentum, where at zero momentum there is an energy gap due to the rest mass of particles \cite{KTrachenko}. In the case of the COCP, the rest energy is directly the energy of natural collective vibrations at the frequency~$\omega_p$.

Another feature of the COCP is the presence of the so-called second high-frequency plasmon peak in the $S(k, \omega)$ spectra, which is found in MD simulations data \cite{Murillo_2019}. It is necessary to note that there is currently no consensus on what causes this effect.

Note that until now there has been no unified theoretical approach that would describe the collective dynamics of the COCP on the wide spatial scale without adjusting parameters. Existing methods either contain adjustable parameters or give satisfactory results in a limited range of wave number values. In this paper, this gap will be filled by use of the self-consistent relaxation theory of collective particle dynamics, which had previously been successfully applied to describe both Yukawa liquids and liquid metals \cite{MFT_PRE, Mokshin}. The main results of this work are as follows. Based on the established correlations between the frequency relaxation parameters characterizing the two-particle, three-particle, and four-particle dynamics of the COCP for different states with the coupling parameter $\Gamma \in [5;100]$ , the expression for the dynamical structure factor $S(k, \omega)$ was derived. It is necessary to note that the obtained expression for $S(k, \omega)$ requires only information about the thermodynamic state and structure of the system as input parameters. The theory correctly reproduces the MD simulations results for $S(k, \omega)$ over a wide range of wave numbers as well as the dispersion laws of the high-frequency plasma mode, decrement of the plasma excitations and frequency of longitudinal plasma excitations.

The paper is organized as follows. In Sec. II, we describe the theoretical formalism related with the self-consistent relaxation theory of ion collective dynamics in the COCP. In Sec. III, the obtained theoretical results are compared with MD simulations data and the results of other theoretical approaches. The main findings are given in the Conclusion (Sec. IV).

\section{Theoretical formalism}
The dynamic structure factor $S(k, \omega)$ is a Fourier transform (in frequency) of the density fluctuations time correlation function $F(k,t)$ known also as the intermediate scattering function \cite{Hansen/McDonald_book_2006, Boon/Yip_1991, Mokshin}:
\begin{equation}
S(k,\omega) = \frac{S(k)}{2\pi}\int_{-\infty}^\infty F(k,t)\exp(\textbf{i}\omega t)dt\,.
\label{Skw_from_Fkt}
\end{equation}
Here, $S(k)$ is the static structure factor; $t$ is the time. For simple liquids in low-$k$ limit (hydrodynamic regime), the exact expression for the dynamic structure factor \cite{Balucani} is known:
\begin{equation}\label{Skw_H}
\begin{gathered}
S^{H}(k, \omega)=\frac{S(k)}{2\pi}\biggl[\frac{\gamma-1}{\gamma}\frac{2D_Tk^2}{\omega^2+(D_Tk^2)^2} \\
+\frac{1}{\gamma}\sum_{j=1}^2\frac{\sigma k^2}{(\omega+(-1)^jc_sk)^2+(\sigma k^2)^2}\biggr].
\end{gathered}
\end{equation}
Here, $\gamma$ is the ratio of the specific heat capacity at constant pressure to the specific heat capacity at constant volume, $D_T$ is the thermal diffusivity coefficient, and $\sigma$ is the sound attenuation coefficient. Equation \eqref{Skw_H} can be derived directly from the linearized Navier-Stokes equations, where the key dynamic variables -- the number density, the energy density and the current -- are treated as slow variables. This equation correctly reproduces the collective dynamics of particles in simple liquids, where the effective interparticle interaction is characterized by a finite length, and it usually provides a phenomenological description of experiments on inelastic light scattering in liquids. A detailed derivation of Eq. \eqref{Skw_H} can be found in the classical monographs  \cite{Balucani, Hansen/McDonald_book_2006}. However, the hydrodynamic theory with Eq. \eqref{Skw_H} for $S(k, \omega)$ is not applicable for the COCP  \cite{MithenRC, MithenPRE2012} that is due to the long-range character of the particle interaction in the COCP. On the other hand, the microscopic theories, which consider the system as an ensemble of interacting particles, turn out to be more efficient. For example, the theory based on the method of frequency moments (FM theory) provides an analytical expression for $S(k, \omega)$, which has no  adjustable parameters \cite{Arkhipov, TkachenkoPRE2020}. Here, the expression for $S(k, \omega)$ is obtained as a result of a fractional-linear transformation of the Nevanlinna function, which has specific mathematical properties and satisfies the sum rules. Further, models based on exponential \cite{Boon/Yip_1991} and Gaussian \cite{Mithen_PRE_2011} memory functions and the model based on a modified Navier-Stokes equation \cite{Murillo_HEDP} are also used to describe the collective dynamics of the COCP. However, these models contain various fitting parameters. This paper presents a theoretical formalism to calculate $S(k, \omega)$ of the COCP, which is based on the self-consistent relaxation theory of collective dynamics in multiparticle systems \cite{Mokshin, AVM_PRE_2001, AVM_JCP, AVM_JPCM, MokshinTMF, MFT_PRE}.

From Eq. \eqref{Skw_from_Fkt}, the following series can be obtained for the function $F(k,t)$:
\begin{equation}
\begin{gathered}
F(k,t) = 1 - \langle\omega^{(2)}(k)\rangle\frac{t^2}{2!}+\langle\omega^{(4)}(k)\rangle\frac{t^4}{4!}+\dots \\
+(-\textbf{i})^l\langle\omega^{(l)}(k)\rangle\frac{t^l}{l!}+\dots \,,
\label{Fkt_series}
\end{gathered}
\end{equation}
where $\langle\omega^{(l)}(k)\rangle$ is the normalized frequency moment $S(k, \omega)$ $l$-th order:
\begin{equation}
\langle\omega^{(l)}(k)\rangle = (-\textbf{i})^l\frac{d^l}{dt^l}F(k,t)\bigg|_{t=0}= \frac{\int_{-\infty}^\infty \omega^l S(k,\omega)d\omega}{{S(k)}}\,.
\label{nmoments_basic}
\end{equation}
From Eq. \eqref{Fkt_series}, one obtains the following expression for the Laplace-transform of the function $F(k,t)$:
\begin{equation}
\begin{gathered}
\widetilde{F}(k, s) = \frac{1}{s} - \frac{\langle\omega^{(2)}(k)\rangle}{s^3} + \frac{\langle\omega^{(4)}(k)\rangle}{s^5} +\dots \\
+ (-\textbf{i})^l\frac{\langle\omega^{(l)}(k)\rangle}{s^{l+1}} +\dots\,.
\label{Fks_series}
\end{gathered}
\end{equation}
On the other hand, the last expression can be rewritten as the continued fraction:
\begin{equation}
\widetilde{F}(k,s) = \cfrac{1}{s+\cfrac{\Delta_1(k)}{s+\cfrac{\Delta_2(k)}{s+\cfrac{\Delta_3(k)}{s+\ddots}}}}
\,.
\label{Fks_cfrac}
\end{equation}
Here, $\Delta_n(k)$ ($n=1, 2, 3 , ...$) is the frequency relaxation parameters, which have a dimension of square frequency \cite{Mokshin, AVM_PRE_2001, AVM_JCP, AVM_JPCM, MokshinTMF, MFT_PRE}. Each of these parameters is related to the corresponding frequency moment of $S(k, \omega)$ via the sum rules \citep{Mokshin}:
\begin{eqnarray}
\Delta _{1}(k) &=&\frac{\langle \omega ^{(2)}(k)\rangle }{\langle \omega
^{(0)}(k)\rangle }, \\
\Delta _{2}(k) &=&\frac{\langle \omega ^{(4)}(k)\rangle }{\langle \omega
^{(2)}(k)\rangle }-\frac{\langle \omega ^{(2)}(k)\rangle }{\langle \omega
^{(0)}(k)\rangle },  \notag \\
\Delta _{3}(k) &=&\frac{\left[ \langle \omega ^{(6)}(k)\rangle \langle
\omega ^{(2)}(k)\rangle -\left( \langle \omega ^{(4)}(k)\rangle \right) ^{2}%
\right] \langle \omega ^{(0)}(k)\rangle }{\langle \omega ^{(4)}(k)\rangle
\langle \omega ^{(2)}(k)\rangle \langle \omega ^{(0)}(k)\rangle -\left(
\langle \omega ^{(2)}(k)\rangle \right) ^{3}},  \notag \\
\,\,\, \dots, \notag \\ 
\Delta _{n}(k) &=& \mathcal{F}\left[ \langle \omega ^{(0)}(k)\rangle, \langle
\omega ^{(2)}(k)\rangle, \dots , \langle \omega ^{(2n)}(k)\rangle \right],  \notag
\end{eqnarray}%
where $\mathcal{F}$ means an algebraic expression.

\begin{figure*}[tbp]
	\centering
\includegraphics[width=15.0cm]{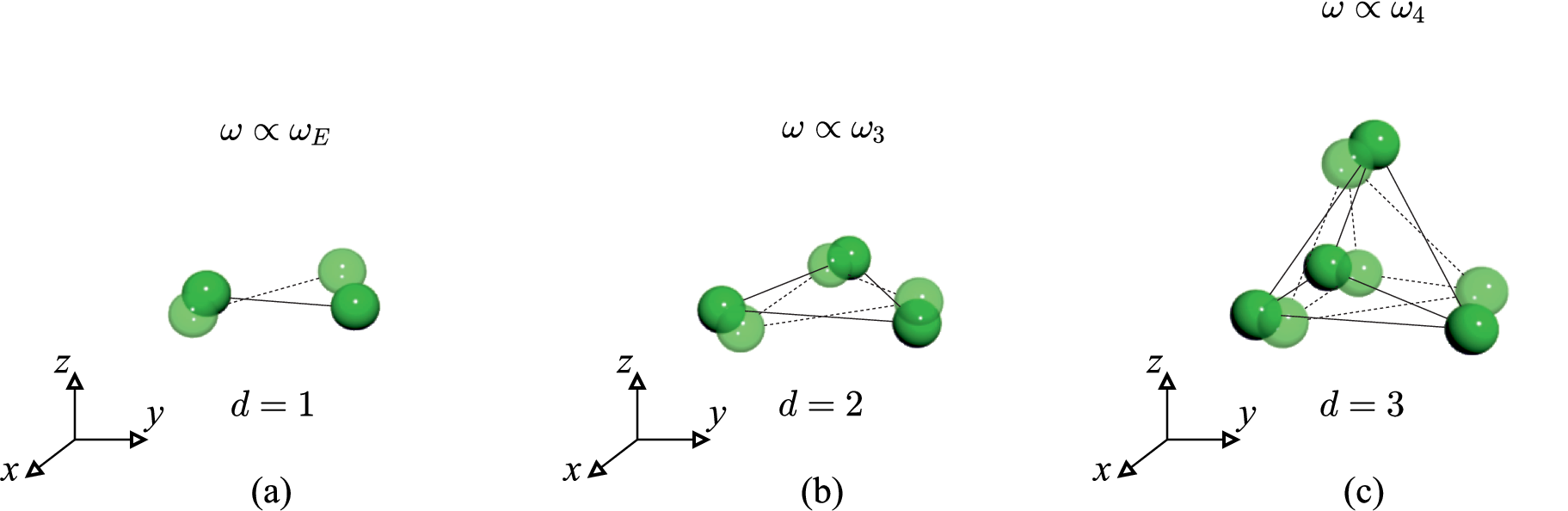}
\caption{Schemes showing arbitrary vibrational $n$-particle groups. The quantity $\omega_n$, where $n=1, 2, 3, ...$, is the average frequency of the corresponding oscillatory circuit, which characterizes the oscillatory dynamics of various $n$-particle groups. Note that in the case of $\omega_2\equiv\omega_E$ we have a linear oscillatory circuit (a), in the case of $\omega_3$ we have a flat oscillatory circuit (b), and in the case of $\omega_4$ we have a three-dimensional oscillatory circuit (c). Particles can oscillate in any direction; however, regardless of this, the dimension of the oscillatory circuit $d$ is preserved.}\label{Part}
\end{figure*}

The following microscopic expressions are known for the frequency relaxation parameters of the first, second, and third orders  \cite{MokshinTMF}:
\begin{subequations}
\begin{eqnarray}
\Delta_1(k) &=& \frac{k_BT}{m}\frac{k^2}{S(k)},  \\ \label{delta_m_2}
\Delta_2(k) &=& 3\left(\omega_E^2 + \frac{k_BT}{m}k^2\right) - \Delta_1(k)\\ \notag
&-& \frac{\rho}{m}\int\nabla_l^2\phi(\textbf{r})\exp(\textbf{i}\textbf{k}\cdot\textbf{r})g(r)d^3\textbf{r}, \\ \label{delta_m_3} 
\Delta_3(k) &=& \frac{\omega_3^4}{\Delta_2(k)} + \Omega_3\left(k\right) . \\ \notag
\end{eqnarray}
\end{subequations}
Here,
\begin{equation}
\omega_E^2 = \frac{\rho}{3m}\int\nabla_l^2\phi(\textbf{r})g(r)d^3\textbf{r}, \nonumber
\end{equation}
is known as the Einstein frequency,
\begin{equation}
\omega_3^4 = \frac{\rho^2}{m^2}\int d^3 \textbf{r}\int d^3 \textbf{r}_1\frac{g_3(\textbf{r},\textbf{r}_1)}{rr_1}\frac{\phi(r)}{dr}\frac{\phi(r_1)}{dr_1}, \nonumber
\end{equation}
is the analog of the Einstein frequency, which characterizes the frequency of vibrational dynamics of different particle triplets, and $\Omega_3(k)$ is the combination integral expressions containing the interaction potential $\phi(r)$, pair distribution function $g(r)$ and three-particle distribution function $g_3(\textbf{r},\textbf{r}_1)$ \cite{MokshinTMF}. In the general case, for the $\Delta_n(k)$, we have the following expression:
\begin{equation}
\label{delta_m_n}
\Delta_n(k) = W\lbrace\omega_n , \Omega_n\left(k \right)\rbrace. 
\end{equation}
Here, $W$ means an algebraic expression, $\omega_n$ is the analog of the Einstein frequency, which characterizes the frequency of vibrational dynamics of different groups of $n$ particles (see Fig. \ref{Part}), and $\Omega_n(k)$ is the combination integral expressions containing the interaction potential $\phi(r)$ and distribution functions beginning from the pair $g(r)$ to the $n$-particle $g_n(\textbf{r},\textbf{r}_1, ... \textbf{r}_{n-2})$ inclusive.Thus, the \textit{n}-th order frequency relaxation parameter $\Delta_n(k)$ is related to the corresponding \textit{n}-particle distribution function of the system and characterizes the vibrational process for various groups of \textit{n} particles \citep{Mokshin, AVM_PRE_2001, AVM_JCP, AVM_JPCM, MokshinTMF, MFT_PRE}.

On the other hand, the quantities $\tau_n(k)=1/\sqrt{\Delta_n(k)}$, where $n=1,2,3,...$, determine the time scale of the corresponding relaxation processes. Thus, the first four quantities in this set -- $\tau_1(k)$, $\tau_2(k)$, $\tau_3(k)$ and $\tau_4(k)$ -- correlate with the time scales of the processes where hydrodynamic variables are exhibited. In turn, these dynamic variables form an orthogonal basis, the first element of which is the density fluctuations \cite{Mokshin, AVM_PRE_2001, AVM_JCP, AVM_JPCM, MokshinTMF, MFT_PRE}. The generalization of the hydrodynamic theory realized within the framework of the self-consistent relaxation theory implies a restriction by the set of frequency relaxation parameters up to and including the fourth order. The time scales of the dynamic variables above the fourth order will be outside the processes that are associated with the structure relaxation.

The key idea of the self-consistent relaxation theory is as follows: beginning from the fourth order the characteristic frequencies of the dynamic variables fluctuations are aligned \citep{Mokshin, AVM_PRE_2001, MokshinTMF, MFT_PRE}, i.e.,
\begin{equation}
\label{Deltas_equal}
\Delta_4(k) = \Delta_5(k) = \Delta_6(k) = ... .
\end{equation}
Using this condition, from Eq. \eqref{Fks_cfrac} one obtains the following analytical expression for $S(k, \omega)$ \cite{Mokshin, AVM_PRE_2001, MokshinTMF, MFT_PRE}:
\begin{eqnarray}\label{Skw_SC}
S(k,\omega) &=& \frac{S(k)}{\pi}\frac{\Delta_1(k)\Delta_2(k)\Delta_3(k)}{\Delta_4(k) - \Delta_3(k)} \\\nonumber
&\times & \frac{\sqrt{\Delta_4(k)}}{\omega^6 + \mathcal{A}_1(k)\omega^4 + \mathcal{A}_2(k)\omega^2 + \mathcal{A}_3(k)}\,,
\end{eqnarray}
where
\begin{eqnarray}
\mathcal{A}_1(k) &=& \frac{\Delta_3^2(k)-\Delta_2(k)[2\Delta_4(k)-\Delta_3(k)]}{\Delta_4(k)-\Delta_3(k)}-2\Delta_1(k)\,, \nonumber
\end{eqnarray}
\begin{equation}
\begin{gathered}
\mathcal{A}_{2}(k) = \frac{\Delta_2^2(k)\Delta_4(k)-2\Delta_1(k)\Delta_3^2(k)}{\Delta_4(k)-\Delta_3(k)} \\
+\frac{\Delta_1(k)\Delta_2(k)[2\Delta_4(k)-\Delta_3(k)]}{\Delta_4(k)-\Delta_3(k)}+\Delta_1^2(k)\,,\notag
\end{gathered}
\end{equation}
\begin{eqnarray}
\mathcal{A}_3(k) &=& \frac{\Delta_1^2(k)\Delta_3^2(k)}{\Delta_4(k)-\Delta_3(k)}\,.\nonumber
\end{eqnarray}
Thus, to calculate $S(k, \omega)$ within the self-consistent relaxation theory, one needs to know the first four frequency relaxation parameters. For the case of the COCP, the following exact microscopic expressions for first- and second-order parameters are known \cite{Mokshin, Arkhipov, Kalman, Golden, Fairushin, Fluids}:
\begin{equation}
\Delta_1(k) = \frac{\omega_p^2(ka)^2}{3\Gamma S(k)} \,,
\label{delta_1_micro}
\end{equation}

\begin{equation}
\begin{gathered}
\Delta_2(k) =  \omega_p^2\left(1+\frac{(ka)^2}{\Gamma} + 2\int_0^{\infty}\frac{j_2(kax)}{x}h(x)dx\right) \\
-\Delta_1(k),
\label{delta_2_micro}
\end{gathered}
\end{equation}
where $x=r/a$ is the dimensionless spatial variable, $j_2(x)$ is the second-order spherical Bessel function, and $h(x)=g(x)-1$. The frequency relaxation parameters $\Delta_3(k)$ and $\Delta_4(k)$ can be determined from MD simulations (details are provided in the Appendix) data through their basic definitions \cite{Mokshin}. Mathematical analysis of these parameters derived from MD simulations  reveals the following correlations:
\begin{subequations}
\label{eq: freq_parameters_approx}
\begin{equation}
\Delta_3(k) \approx \frac{3}{2}\Delta_2(k) + \omega_0^2, 
\label{approx1}
\end{equation}
\begin{equation}
\Delta_4(k) \approx \frac{4}{3}\Delta_3(k) + \omega_1^2(k) \approx 2\Delta_2(k) + \frac{4}{3}\omega_0^2 + \omega_1^2(k),
\label{approx2}
\end{equation}
\end{subequations}
where
\begin{equation}
\omega_0^2 = \frac{3\,\omega_p^2}{\sqrt{\Gamma}}; \,\, \omega_1^2(k) = \frac{\,\omega_p^2\sqrt{\Gamma}}{7ka}.
\nonumber
\end{equation}

\begin{figure}[h]
\centering
\includegraphics[width=8.6cm]{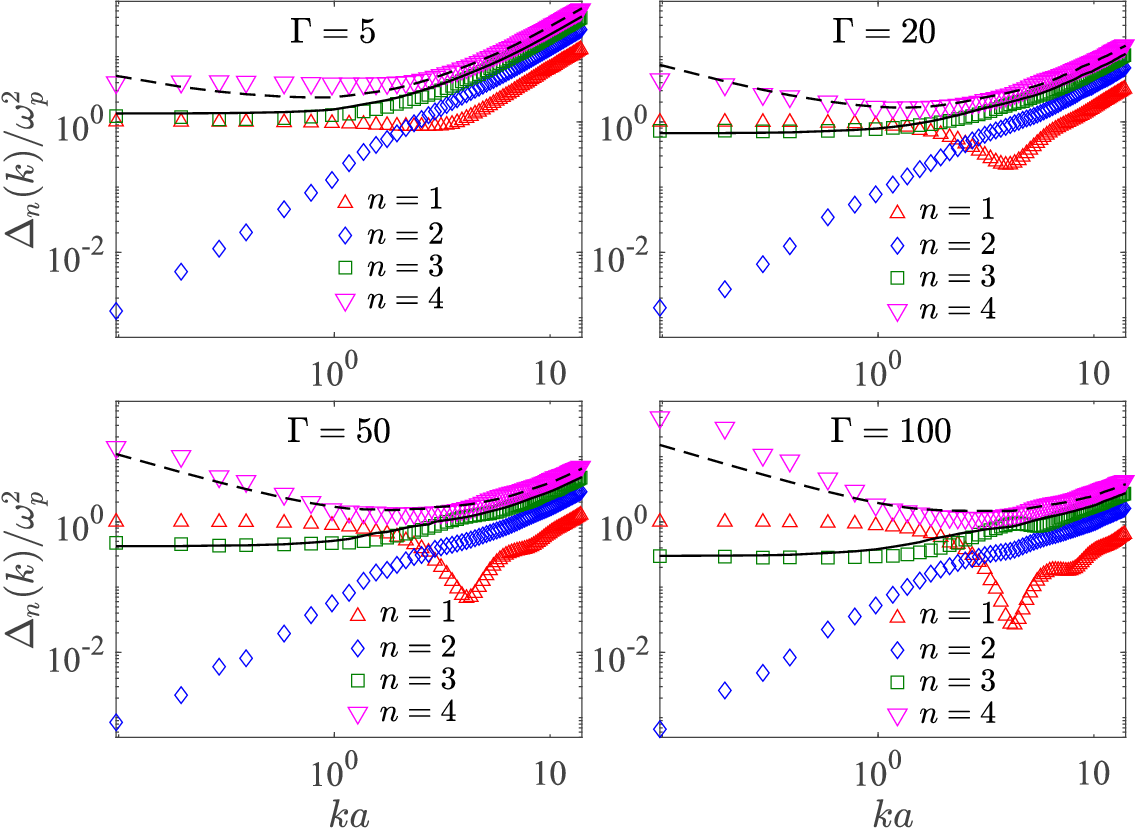}
\caption{Dispersion dependences of the frequency relaxation parameters reduced to $\omega_p^2$ [symbols -- calculations based on simulation data; solid and dashed lines -- calculations using approximate correlation relations (\ref{approx1}) and (\ref{approx2}), respectively].}\label{disp_deltas}
\end{figure}

It is necessary to note that the similar correlations were obtained for the Yukawa plasmas one-component (YOCP) \cite{MFT_PRE}. In contrast to this case, Eq. \eqref{approx2} for $\Delta_4(k)$ contains the $k$ dependent term $\omega_1^2(k)$, which occurs due to the long-range nature of the Coulomb interaction potential \eqref{coulomb_pot}. Figure \ref{disp_deltas} shows $k$ dependencies of the reduced frequency relaxation parameters for the COCP at various $\Gamma$. As can be seen, Eqs. \eqref{approx1} and (\ref{approx2}) satisfactorily reproduce well the MD simulations results for $\Delta_3(k)$ and $\Delta_4(k)$. In fact, relations \eqref{approx1} and \eqref{approx2} indicate a correspondence between two-particle correlations and three- and four-particle correlations. The theoretical model presented in this paper is obtained within the framework of the self-consistent relaxation theory \cite{MFT_PRE, Mokshin, AVM_PRE_2001, AVM_JCP, AVM_JPCM, MokshinTMF}, which is modified for the case of the COCP, where the frequency relaxation parameters are related to each other according to certain correlation relations \eqref{approx1} and \eqref{approx2}. These relations represent an empirical result: as follows from molecular dynamics simulation data, these relations are satisfied for the thermodynamic states where the COCP is a fluid-like system (i.e. at $\Gamma \in [5;100]$). The theoretical model presented applies directly to this region of states.

\begin{figure*}[tbp]
\begin{center}
\includegraphics[keepaspectratio,width=\linewidth]{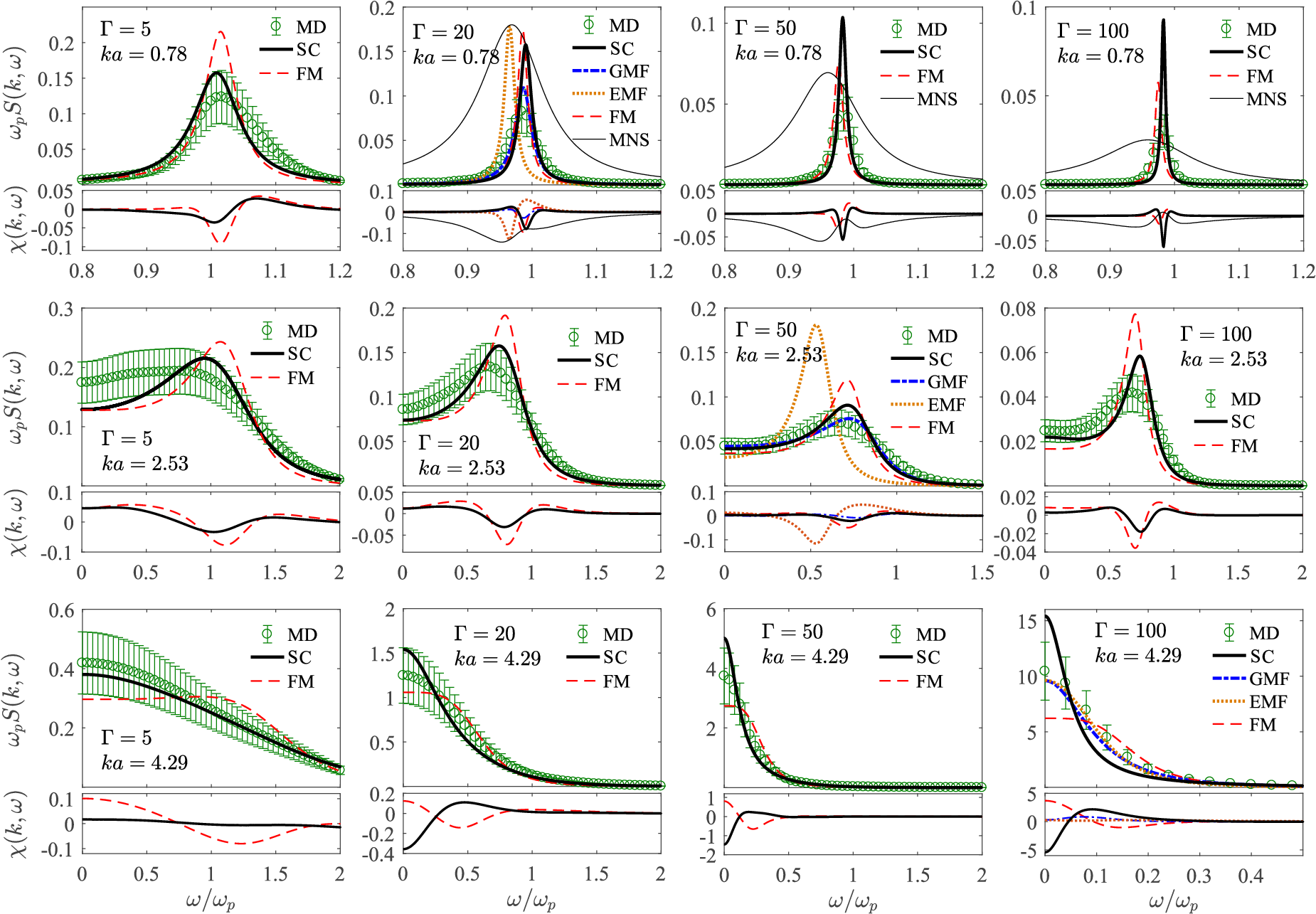}
\end{center}
\caption{Top panels: spectra of $S(k, \omega)$ multiplied by the plasma frequency at different values of the coupling parameter $\Gamma$. Here theoretical results from Eq. \eqref{Skw_SC} shown by black solid lines are compared with MD simulations data given by green circles, with results of the FM theory \cite{Arkhipov} given by red dashed lines, with results of the exponential memory function model (EMF) given by brown dashed lines \cite{Boon/Yip_1991}, results of the Gaussian memory function model (GMF) \cite{Mithen_PRE_2011} given by blue dashed lines and results of the model based on and modified Navier-Stokes equation (MNS) \cite{Murillo_HEDP} given by a thin black line. Bottom panels: differences between the simulation data and the corresponding theoretical values.}\label{DSF}
\end{figure*}

Analysis of expression \eqref{Skw_SC}  allows one to obtain the dispersion equation for the high-frequency plasma mode:
\begin{equation}
s^{3}+\mathcal{B}_1(k)s^{2} +\mathcal{B}_2(k)s+\mathcal{B}_1(k)\Delta_1(k)=0,  \\
\end{equation}
where
\begin{equation}
\mathcal{B}_1(k) = \frac{2\Delta_1(k)\sqrt{\Delta_4(k)}}{2\Delta_4(k)-\Delta_3(k)}, \nonumber
\end{equation}
\begin{equation}
\mathcal{B}_2(k) = \Delta_1(k)+\mathcal{B}_1(k)\sqrt{\Delta_4(k)}. \nonumber
\end{equation}
Solution of this equation yields $s(k)=\pm i\omega _{c}(k)-\delta (k)$ with dispersion for the side peak of $S(k, \omega)$:
\begin{equation}
\omega _{c}(k)=\sqrt{3}\left( \sqrt[3]{Z(k)-q(k)}+\sqrt[3]{Z(k)+q(k)}\right),  \label{eq: omega_c}
\end{equation}%
and the decrement dispersion of plasma excitations,
\begin{equation}
\delta (k)=\sqrt[3]{Z(k)+q(k)}-\sqrt[3]{Z(k)-q(k)}-\frac{\mathcal{B}_2(k)}{3},
\label{eq: plasm_att}
\end{equation}%
where
\begin{eqnarray}
Z(k) &=&\sqrt{p^{3}(k)+q^{2}(k)},  \notag \\
p(k) &=&\frac{\mathcal{B}_2(k)}{3} + \left(\frac{\mathcal{B}_1(k)}{3}\right)^2,  \notag \\
q(k) &=&\frac{\mathcal{B}_1(k)}{54}\left(2\mathcal{B}_1(k)^2-9\mathcal{B}_1(k)\sqrt{\Delta_4(k)}+18\Delta_1(k)\right) .  \notag
\end{eqnarray}

Obviously, by analogy with the hydrodynamic expression \eqref{Skw_H}, the values $\omega_c(k)$ and $\delta(k)$ defined by formulas \eqref{eq: omega_c} and \eqref{eq: plasm_att}, respectively, will characterize the position and width of the side peak in $S(k, \omega)$ spectra.

\section{Results and discussion}
The theoretical results are compared with the data of MD simulations and with the results of other models, in particular, with the FM theory. In the work of \cite{Arkhipov}, based on physical considerations, an expression for $S(k, \omega)$ was found that is similar to Eq. (\ref{Skw_SC}). A detailed discussion of the correspondences between the self-consistent relaxation theory and the method of frequency moments was given in Ref. \cite{MFT_PRE}.

Figure \ref{DSF} shows the $S(k,\omega)$ spectra of the COCP for different dimensionless wave numbers $ka$ and coupling parameters $\Gamma = 5, 20, 50$ and $100$. These $\Gamma$ values correspond to the liquid phase of the COCP. It can be seen that for the considered values of the coupling parameter $\Gamma$ and wave number $k$, the self-consistent relaxation theory reproduces the results of MD simulations quite accurately and describes all the features of these spectra. At small wave numbers $k<k_m/2$, where $k_m$ is the wave number corresponding to the first maximum in the static structure factor $S(k)$, the spectra $S(k,\omega)$ of the COCP, as expected, contain only the high frequency components at the frequencies near to $\omega_p$. As the dimensionless wave number $ka$ increases, beginning from values comparable to $k_m/2$, the zero component appears and the high-frequency component disappears. This feature is characteristic of all classical simple liquids with a short-range interparticle interaction potential. This means that beginning from the wave numbers $k=k_m/2$ and higher, i.e. on spatial scales that correspond to several mean interparticle distances, the long-range character of the Coulomb interaction ceases to play an appreciable role in the particle dynamics. Note that Eq. (\ref{Skw_SC}) in some cases gives somewhat better agreement with the results of MD simulations than the FM theory \cite{Arkhipov}, as well as models based on the exponential memory function \cite{Boon/Yip_1991} and modified Navier-Stokes equation \cite{Murillo_HEDP}. The model based on the Gaussian memory function \cite{Mithen_PRE_2011} gives good agreement with MD simulations data, but it contains a fitting parameter -- the so-called relaxation time.

\begin{figure*}[tbp]
	\centering
\includegraphics[width=17.0cm]{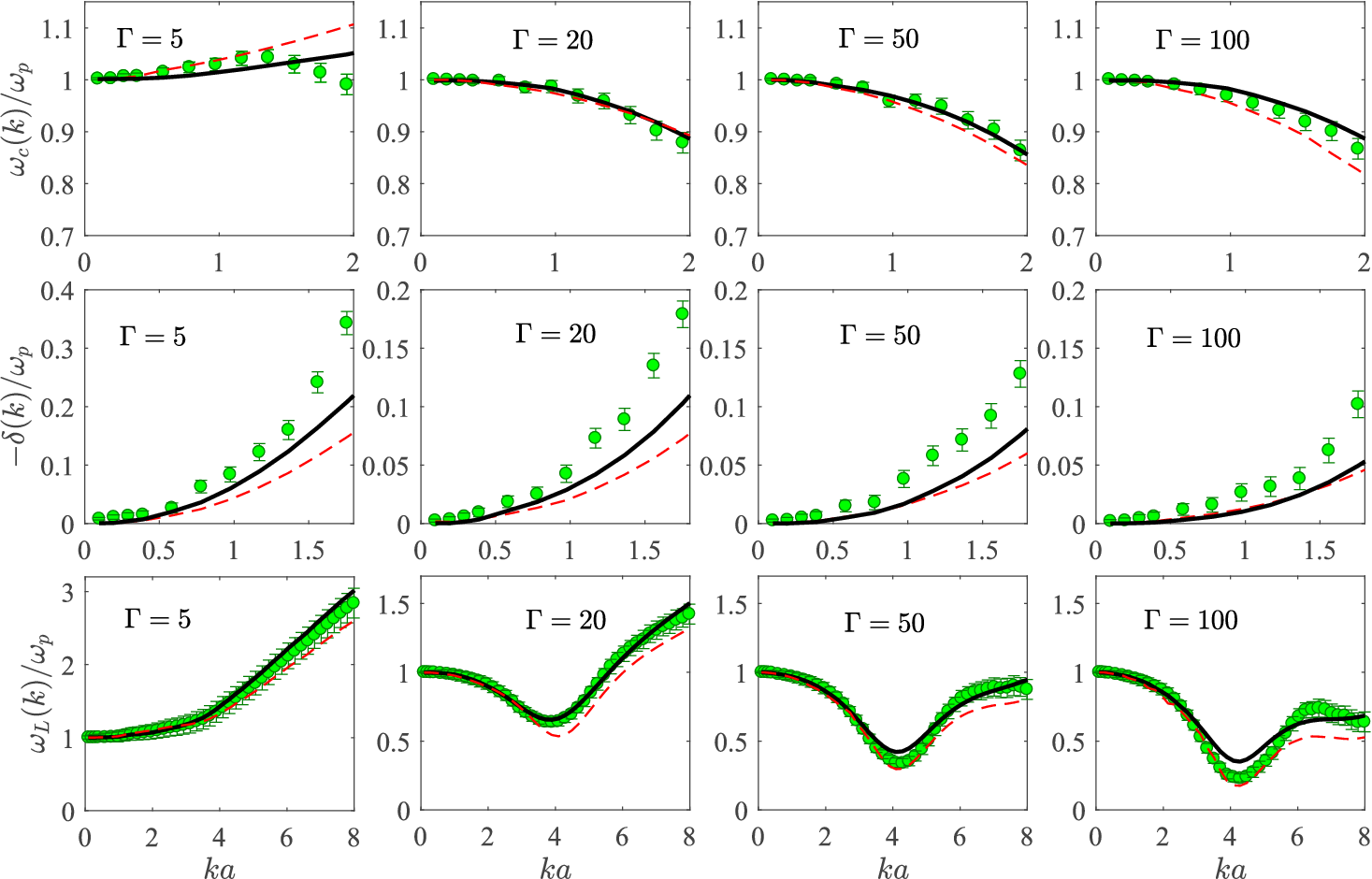}
\caption{Wave number dependencies of the frequency $\omega_c(k)$ (top row), decrement of plasma excitations $\delta(k)$ (middle row) and longitudinal plasma excitations $\omega_L(k)$ (bottom row) plotted at different values of the coupling parameter $\Gamma$. The black solid lines represent theoretical results obtained using expressions (\ref{eq: omega_c}), (\ref{eq: plasm_att}), and (\ref{eq: omega_L}), the red dashed lines show the results of the  moment theory \cite{Arkhipov}, and the green circles show the MD simulations data.}\label{Disp}
\end{figure*}

To obtain the $k$ dependence of the longitudinal collective excitations frequency $\omega_L$, we consider the spectral density of the longitudinal current correlation function $C_L(k, \omega)$, which is directly related to $S(k, \omega)$ as
\begin{equation}
C_{L}(k,\omega )=\frac{3\Gamma\omega^2}{(\omega_pka)^2}S(k,\omega ).
\label{eq: CL}
\end{equation}
Using this relation, from Eq. \eqref{Skw_SC} one can obtain the analytical expression for the dispersion law $\omega _{L}(k)$ of longitudinal plasma excitations:
\begin{equation}
\omega _{L}(k)=\sqrt{C_{+}(k)+C_{-}(k)-\frac{\mathcal{A}_{1}(k)}{6}},
\label{eq: omega_L}
\end{equation}%
where
\begin{equation}
C_{\pm }(k)=\sqrt[3]{\frac{\mathcal{A}_{3}(k)}{4}-\frac{\mathcal{A}_{1}^{3}(k)}{216}\pm \sqrt{%
\frac{\mathcal{A}_{3}^{2}(k)}{16}-\frac{\mathcal{A}_{3}(k)\mathcal{A}_{1}^{3}(k)}{432}}}.  \notag
\end{equation}%
Figure \ref{Disp} presents the dispersion characteristics of the COCP. It can be seen (top and middle rows) that Eqs. (\ref{eq: omega_c}) and (\ref{eq: plasm_att}) reproduce the MD simulations results for the dispersion characteristics $\omega_c(k)$ and $\delta(k)$ very well. From the bottom row, it is clear that Eq. (\ref{eq: omega_L}) enables one to correctly calculate $\omega_L(k)$ over a wide range of the COCP parameter changes. The proposed theoretical formalism correctly reproduces asymptotes of dispersion dependencies for low wave numbers and the so-called roton minima \cite{Arkhipov} at the COCP states with $\Gamma = 5, 20, 50$ and $100$.

A remarkable fact is that both the self-consistent relaxation theory and the FM theory produce expressions for the characteristics of the collective particle dynamics in terms of frequency moments and/or frequency relaxation parameters. In addition, both theories are consistent with each other: one can formulate a condition for high order frequency relaxation parameters under which self-consistent relaxation theory produces FM theory results. This point is discussed in detail in Ref. \cite{MFT_PRE} (see Supplemental Material). A characteristic feature of FM theory is that it is based on the Nevanlinna parameter function. The most important advantage of this theory is that the obtained analytical expressions for the dynamical structure factor and other quantities of collective particle dynamics do not contain any fitting parameters \cite{Arkhipov, TkachenkoPRE2020} and also the theory does not rely directly on any empirical results leading to expressions similar to correlation relations \eqref{approx1} and \eqref{approx2}. On the other hand, a feature of FM theory is that it does not take into account the manifestation of an independent central (Rayleigh) component in the spectra of the dynamic structure factor at wave numbers comparable with $ka=4.29$ and higher (see Fig. \ref{DSF}).

As mentioned above, states close to an ideal Coulomb gas with $\Gamma\ll1$ have a positive dispersion of high-frequency collective excitations [see Eq.\eqref{BGdisp}], while a Coulomb (Wigner) crystal with $\Gamma\gtrsim175$ is characterized by a negative dispersion of these excitations \cite{Baus}. Thus one can expect that the state with $\Gamma_c\approx 9.5$, which is the boundary for regimes with positive and negative dispersions, will correspond to a crossover between regimes with \textit{gas like} and \textit{solid like} collective ion dynamics. On the other hand, the disappearance of the so-called roton minima observed at $\Gamma_c$ represents one of the conditions of the Frenkel line \cite{Brazhkin3}, which in the phase diagram of an arbitrary system separates the thermodynamic states with \textit{gas like} and \textit{solid like} particle dynamics \cite{Brazhkin3, Brazhkin1, Brazhkin2, Trachenko}. Consequently, in the case of the COCP, there is a direct correspondence between the known value of $\Gamma_c$ and the Frenkel line, which will be located at $\Gamma_c\approx 9.5$ in the phase diagram.

To determine $\Gamma_c$ in the framework of the proposed formalism, it is necessary to find the approximation of Eq. \eqref{eq: omega_L} at small $k$ as a quadratic polynomial of the form
\begin{equation}
\omega_L^{(\textrm{lk})}(k) \approx \omega_p\left(1+\alpha(ka)^2\right).
\end{equation}
The positive values of the coefficient $\alpha$ correspond to positive dispersion, whereas the negative values correspond to negative dispersion. Table I shows the values of this coefficient for various $\Gamma$.
\begin{table}[h]\caption{Values of the coefficient $\alpha$ at various $\Gamma$.}
\begin{center}
\begin{tabular}{l|c|c|c|c}
$\Gamma$\, & 5 & 20 & 50 & 100 \\
 \hline
 $\alpha$ &\, 0.021 \,&\, -0.020 \,& \,-0.029 \,& \,-0.031 
\end{tabular}
\end{center}
\end{table}
Using three alpha values corresponding to $\Gamma = 5$, $20$, and $50$, one can construct an approximation of the dependence $\alpha(\Gamma)$ in the following view:
\begin{equation}
\alpha(\Gamma)=2.237\cdot10^{-4}\Gamma^2 - 0.017\Gamma + 0.165.
\label{alpha_pol}
\end{equation}
$\Gamma_c$ will correspond to the condition with $\alpha(\Gamma)=0$. Then, we find $\Gamma_c\approx 11.42$. As can be seen, this result is close to $\Gamma_c\approx9.5$ obtained from the large-scale MD simulations in Ref. \cite{MithenAIP}.

\section{Conclusion}
Thus, in this paper, correlation ratios between the frequency relaxation parameters characterizing the three- and four-particle dynamics with the parameters characterizing the two-particle dynamics were obtained for the case of the COCP at the values of the coupling parameter $\Gamma=5$, $20$, $50$ and $100$. The application of the obtained correlations enables one to describe all the features of this non-trivial multiparticle system within the self-consistent relaxation theory without any fitting parameters. In spite of the fact that in the realized approach all correlations are reduced to pairs correlations, it turns out to be sufficient to describe a system of particles with the long-range Coulomb interactions. The calculated dynamic structure factor and dispersion characteristics are consistent with molecular dynamic simulation data. The obtained discrepancies between the theoretical results and the MD simulations data are comparable with those given by the theory based on the frequency moments method.

\section{ACKNOWLEDGEMENTS} This work was supported by the Russian Science Foundation (Project No. 19-12-00022). The authors are grateful to I. M. Tkachenko and S. A. Khrapak for helpful discussions.

\section{APPENDIX: MOLECULAR DYNAMICS SIMULATION DETAILS} MD simulations of the COCP were performed in the LAMMPS package \cite{LAMMPS} for the equilibrium configuration of the COCP at $\Gamma = 5, 20, 50$ and $100$ in the NVT ensemble. The simulation cell contained 64,000 particles interacting through the Coulomb potential. Periodic boundary conditions in all directions were applied to the cell and the PPPM fast summation method was used. The equations of motion of the particles were integrated using the velocity-based Verle algorithm with a time integration step $\tau=0.01/\omega_p$.

\end{document}